\documentclass[final,3p,times]{elsarticle}

\usepackage{graphics}
\usepackage{graphicx}
\usepackage{amssymb}
\usepackage{amsthm,latexsym}
\usepackage{lineno}
\usepackage{color}
\usepackage{multicol}

\biboptions{compress}
\journal{Physics of the Dark Universe}
\begin{document}
\begin{frontmatter}


\title{Solar WIMPs Unraveled: Experiments, astrophysical uncertainties, and interactive Tools}

\author[label1]{Matthias Danninger}
\address[label1]{Department of Physics, University of British Columbia, Vancouver BC, Canada}
\ead{matthias.danninger@cern.ch}
\author[label2]{Carsten Rott}
\address[label2]{Department of Physics, Sungkyunkwan University, Suwon 440-746, Korea}
\ead{rott@skku.edu}

\begin{abstract}
The absence of a neutrino flux from self-annihilating dark matter captured in the Sun has tightly constrained some leading particle dark matter scenarios. The impact of astrophysical uncertainties on the capture process of dark matter in the Sun and hence also the derived constraints by neutrino telescopes need to be taken into account. In this review we have explored relevant uncertainties in solar WIMP searches, summarized results from leading experiments, and provided an outlook into upcoming searches and future experiments. We have created an interactive plotting tool that allows the user to view current limits and projected sensitivities of major experiments under changing astrophysical conditions. 
\end{abstract}

\begin{keyword}
Dark Matter \sep Indirect detection \sep Neutrino Astronomy
\end{keyword}

\end{frontmatter}


\section{Introduction}

The presence of dark matter (DM) in the universe has been inferred through its gravitational interactions. Beyond this strong observational evidence, very little is known about the nature of DM. One of the most promising and experimentally accessible candidates for DM are so-called Weakly Interacting Massive Particles (WIMPs)~\cite{Bertone:2004pz}. WIMP like particles arise naturally in many extensions of the Standard Model of particle physics (SM), and are predicted to have a mass in the range of a few GeV to a few TeV~\cite{SUSY,UED,cheng}.

Despite several widely discussed observations which hint at possible DM signals~\cite{DAMASavage, COGENTmodulation, CDMSsilicon}, no undisputed direct experimental evidence for WIMPs exists. Experimental efforts are divided into three main techniques. Direct detection experiments look for a nuclear recoil signal within the detector volume from weak-scale scattering of WIMPs with target nuclei. Indirect detection experiments aim to detect primary or secondary particles created in WIMP pair-annihilations or decays, such as photons, neutrinos and antimatter. Accelerator searches aim to find DM through its production in particle collisions. These search strategies are complementary and have reached comparable sensitivities to viable DM models~\cite{silverwwod, Baer:2009bu, Buchmueller:2012hv,Edsjo:2004pf}. 

Neutrino telescopes like IceCube~\cite{icecube}, Super-Kamiokande~\cite{superk-original}, ANTARES~\cite{antares}, and Baksan~\cite{Baksan} search indirectly for DM via neutrinos from DM self-annihilations.
These telescopes have a unique discovery potential for DM by searching for a high-energy neutrino flux from the Sun. Such a striking signature is predicted, as DM may be captured in large celestial bodies like the Sun where self-annihilation to SM particles can result in a flux of high-energy neutrinos. This indirect search is sensitive to the cross-section for WIMP-nucleon scattering, which initiates the capture process in the Sun. Solar WIMP searches present a special case of indirect DM searches, as they are in general not sensitive to the self-annihilation cross-section but only benefit from the self-annihilating nature of WIMPs to test their scattering cross-section with matter, as in direct detection experiments. Limits on WIMP-nucleon scattering cross-sections from neutrino telescopes are very stringent and depend only weakly on Astrophysical assumptions, e.g. WIMP velocity distributions. Following earlier work~\cite{Choi:2013eda,Rott:2011fh,Bruch:2008rx,Peter:2009ak,Ellis:2008hf,Bottino:1999ei}, we developed an interactive plotting tool\footnote{Preliminary link to interactive tool (https://mdanning.web.cern.ch/mdanning/public/Interactive\_figures/), including a convenient bash script for download and testing.}, which incorporates the dominant sources of astrophysical uncertainties and current experimental constraints. This comprehensive interactive tool aims to quickly visualize the impact of various uncertainties on experimental limits.

This paper is structured as follows: In section~\ref{theory}, we first review details about the WIMP capture process in the Sun. Sections~\ref{analysis_description} and~\ref{current_results} respectively give an overview of analysis strategies and most recent results for solar DM searches with neutrino detectors. In section~\ref{astro_uncert}, we discuss different sources of Astrophysical uncertainties that are important for these searches and included in the interactive plot. 
In the final section~\ref{conclusion} we give an outlook on future prospects and conclude our results. \ref{toolDescription} provides a technical description of the interactive tool and explains available input options.


\section{Solar WIMPs}\label{theory}

WIMPs from the Milky Way DM halo could be gravitationally captured by the Sun and accumulate in its centre. We here summarise the standard capture calculation and refer the interested reader to Ref.~\cite{Gould:1987ir,Gould:1991hx,Wikstrom:2009kw} for more detailed reviews. WIMP capture is initiated by an elastic scattering process in which a WIMP could loose enough energy to fall below the escape velocity of the Sun, and hence becomes gravitationally bound to it. In subsequent scatters the WIMP can loose more energy and eventually sink to the centre of the gravitational well and thermalize. 

The differential WIMP capture rate in the Sun for a WIMP of mass $m_{\chi}$ can be obtained by dividing the Sun into shell volumes $dV$ at a distance $r$ from its centre and computing the capture rate on nucleus $i$ for each shell. The total capture rate by the Sun, $C$, can be obtained by integration over the shell volumes up to the radius of the Sun $R_{\odot}$:
\begin{equation}
C = \int_0^{R_{\odot}} 4\pi r^2 dr \sum_{i} \frac{dC_{i}(r)}{dV},
\end{equation}
with the differential capture rate is given by~\cite{Gould:1987ir}
\begin{equation}\label{eq:capture}
\frac{dC_{i}}{dV}= \int_{0}^{u_{max}}  du \int d\Omega_{w} f(u) u w^2 \sigma_{i} n_{i} \frac{\rho_{\chi}}{m_{\chi}},
\end{equation}
$f(u)$ is the velocity distribution function of the WIMPs in the Solar reference frame normalised to unity and $\rho_{\chi}$ the dark matter density.  
$\sigma_{i}$ is the elastic-scattering cross-section at zero-momentum transfer of a WIMP with nucleus $i$ with mass, $M_i$, and $n_{i}$ its number density of nuclear species $i$ at the corresponding shell volume. $w$ is the velocity at a given shell, which is related to the escape velocity $v_{esc}$ at the shell and WIMP velocity at infinity,$u$ , by $w = \sqrt{u^2 + v_{esc}^2}$.

The integration upper limit $u_{max}$ ensures that only WIMPs that can scatter to a velocity below the escape 
velocity, $v_{esc}$, are included. The upper limit is given by: 
\begin{equation}
u_{max} = 2 \frac{\sqrt{M_i m_{\chi}}}{m_{\chi} - M_{i}} v_{esc}.
\end{equation}

For the simplest case of a Maxwellian velocity distribution, f(u) has the following form:
\begin{equation}
 f(u) = \sqrt{\frac{3}{2\pi}} \frac{u}{v_{\odot}v_{rms}} 
\Bigl( exp \Bigl(- \frac{3(u - v_{\odot})^2}{2 v_{rms}^2} \Bigr) - 
exp \Bigl( - \frac{3(u + v_{\odot})^2}{2 v_{rms}^2} \Bigr) \Bigr).
\end{equation}
here $v_{\odot}$ is the circular velocity of the Sun and $v_{rms}$ the local velocity dispersion of the dark matter halo.

As a result of the continuous DM capture by the Sun, we could find ourselves in the vicinity of a very dense DM accumulation that exceeds the average local DM density by orders of magnitude. While the DM accumulation in the Sun is completely insignificant to the mass of the Sun~\cite{Edsjo:2010bm}, the overdensity at the Sun's centre could result in DM annihilations at significant rates. The number density of DM in the Sun, $N$, is given by: $dN/dt = C - C_{A} N^2$, where $C$ is the dark matter capture rate and $C_{A}$ describes DM annihilation. $C_{A}$ depends on the thermally averaged product of the total annihilation cross-section and the relative particle velocity per volume after capture. 
The DM capture rate in the Sun ($C$) can be considered constant over time as the parameters effecting it are not expected to change. Capture depends on the DM density, relative velocity of WIMPs and Sun, scattering cross-section, and solar composition. The annihilation rate will steadily increase up to a point where as much DM is annihilated as is captured, which is known as the equilibrium condition ($dN/dt=0$). Equilibrium is typically reached at time scales that are roughly two orders of magnitude smaller than the age of the solar system~\footnote{For this value we assume a DM self-annihilation cross-section of the size of the thermal relic cross-section; However, the time scale strongly depends on the DM scenario and should be evaluated for each case~\cite{Rott:2011fh}.}. Hence, if equilibrium between capture and annihilation has been achieved, the DM annihilation rate will only depend on the total scattering cross-section. For DM models with large self-annihilation cross-sections equilibrium will be reached faster than for models with smaller self-annihilation cross-sections. The equilibration time, $\tau$, determines how fast equilibrium is reached and the annihilation rate,$\Gamma_{A}$ , in the Sun can be described by:
\begin{equation}
\Gamma_{A} (t) = \frac{C}{2}\tanh^2\left( \frac{t}{\tau}\right),
\end{equation}
where $t$ = $t_{\odot}$ is the age of the Sun and the equilibration time given by $\tau = 1 / \sqrt{C \times C_{A}}$. 

WIMPs captured in the Sun could also become unbound in a process known as evaporation, which is relevant for WIMP masses below 4\,GeV and not focus of our review. It has hence been ignored. Detailed studies of the capture and annihilation process for WIMP masses below 4\,GeV are discussed elsewhere~\cite{Griest:1986yu,Gould:1987ju,1990ApJ...356..302G,Bernal:2012qh,Kappl:2011kz}.

The WIMP model dependent interaction cross-section is composed of the spin-independent component (SI) and the spin-dependent component (SD) of the interaction cross-section ($\sigma$). 
As the Sun is primarily a proton target, it could capture DM very effectively via SD scattering, where contributions from heavier elements can be ignored. This is different for capture via SI scattering, where it is important to sum over all elements in the Sun, owing to $\sigma_{ \mathrm{SI}} \sim \mathrm{A}^2$, where A is the atomic mass number. As a result, the SI cross-section depends on detailed information on the solar abundance of elements.

The final states in a DM annihilation are model specific and depending on the theory a mix of various final states can be produced. Neutrinos that can escape the Sun and mark an observable WIMP signal can be produced directly in the annihilation or through decays of annihilation products. Light quarks ($u$, $d$, $s$) hadronize quickly to form mesons that have long enough lifetimes to interact with the solar medium before they can decay or in the case of neutral pions decay without producing neutrinos. The WIMP annihilation products, which produce energetic neutrinos, are $c$, $b$, and $t$ quarks, $\tau$-leptons, and gauge bosons. Neutrinos from decays of short lived $c$, $b$, and $t$ quarks have a soft spectrum (lower energy), as the quarks initially loose energy during hadronization and often complex decay chains. WIMP annihilations into ${\rm W}^{+}{\rm W}^{-}$ result in a hard (higher energy) spectrum through the secondary direct decay into charged leptons and neutrinos. Below $m_{\rm W}$, the annihilation into $\tau^{+}\tau^{-}$ is assumed as the channel producing the highest energy neutrinos. Two end points of the spectrum are chosen to approximately bracket the range of all models: the soft $b\bar{b}$ and hard ${\rm W}^{+}{\rm W}^{-}$ ( $\tau^{+}\tau^{-}$ below $m_{\rm W}$) channels (each with 100\% branching).

The neutrinos produced as part of the DM annihilation process in the Sun could be observable at Earth by various neutrino telescopes and detectors. To predict the neutrino signal at Earth, neutrino propagation from the solar centre, where annihilations take place, to its surface and then from the Sun to Earth have to be considered. The DarkSUSY~\cite{Gondolo:2004sc} and WimpSim~\cite{Blennow:2007tw} packages for numerical calculation of interactions and oscillations in a fully three flavour scenario, including  regeneration from tau leptons are utilized to predict neutrino fluxes at the detector.
The Sun becomes optically thick to neutrinos at about 1~TeV and already at 100~GeV some significant absorption is expected. Additionally, DarkSUSY is used to convert limits on the neutrino induced muon flux~\cite{Wikstrom:2009kw} or neutrino flux~\cite{Rott:2011fh} from the Sun to limits on the SD or SI WIMP-nucleon scattering cross-section. 


\section{Analysis strategies for solar WIMP signals in neutrino detectors}\label{analysis_description}

In this section we discuss data analysis strategies used in searches for solar WIMP signals in neutrino detectors. We give an overview over sources of background and provide a description of commonly used background rejection techniques.

The main background in a search for DM annihilations in the Sun consists of muons and neutrinos created in cosmic ray (CR) interactions in the Earth's atmosphere. This steady flux of muons and neutrinos  dominates at depths relevant for neutrino telescopes~\cite{Nakamura:2010zzi}. 
The leading production of muons is via the leptonic or semi-leptonic decays of charged pions or kaons.
The predominant trigger rate of neutrino telescopes originates from atmospheric muons.
The zenith angle range is constrained to angles of less than $85^{\circ}$, as atmospheric muons get eventually absorbed in the Earth.

While the atmospheric muon flux can be rejected relatively easy due to topological and directional selection criteria, atmospheric neutrinos present a continuous and partially irreducible background to solar WIMP signals. The atmospheric neutrino background is created in decays of secondary particles in cosmic-ray air showers and hence present from all directions. The atmospheric neutrino energy spectrum is characterized by two components, a conventional flux in the GeV-TeV range~\cite{FluxOfAtmoNu} from pion and decay decays and a prompt component from charmed mesons.
The flux of prompt atmospheric neutrinos, predicted from the semi-leptonic decay of charmed particles~\cite{PromptComponent} is expected to be relevant at high energies ($\mathcal{O}(100)$ TeV).
The cross-sections for their production are small and have not yet been experimentally identified~\cite{IC59Diffuse}, and therefore can be neglected at energies relevant for solar WIMP searches.  

A further component, which potentially may fake a DM signal, arises from CR interactions in the solar atmosphere. 
As the solar atmosphere is less dense at typical interaction heights than the Earth's atmosphere, a larger fraction of mesons will decay instead of interacting, enhancing the $\nu$-flux component from the solid angle of the solar disc~\cite{solarAtmNu}. The number of events expected from this background at final analysis level were calculated for the latest IceCube results~\cite{Aartsen:2012kia} in Ref.~\cite{MDthesis}, using predicted parameterizations of the solar atmospheric neutrino fluxes~\cite{solarAtmNu,solarAtmNu2, osci-solarAtmNu}. Events corresponding to the highest flux model are expected to make up less than 3\% of the total expected atmospheric neutrino background and further reduced when considering neutrino oscillations~\cite{solarAtmNu2, osci-solarAtmNu}.\\

To detect high-energy neutrinos in statistical significant numbers, a combination of large detector and long observation times are needed. Neutrino telescopes combined large volumes with a nearly 100\% duty cycle making them ideal to search for neutrinos from DM annihilations.
We can differentiate between two detection technologies, water Cherenkov and muon counters. The latter technology is used in the Baksan underground scintillator telescope~\cite{Baksan}. Water or ice Cherenkov detectors, like IceCube~\cite{icecube}, Super-Kamiokande~\cite{superk-original} (Super-K), and ANTARES~\cite{antares}, detect Cherenkov light radiated by charged particles that are produced in interactions with nuclei inside or close by the detector.

The analysis strategy in a search for solar WIMP signals is common for all neutrino telescopes. Based on distributions of event multiplicities and observables from signal simulations and experimental data, selection criteria are placed to reduce the content of atmospheric muon events. This is repeated until the event selection is dominated by signal-like atmospheric neutrino events. The final step is a directional search comparing the observed number of events from the direction of the Sun with the background-only hypothesis. 

For most of the recent results~\cite{SuperKWIMPlatest, Adrian-Martinez:2013ayv, Boliev:2013ai}, analyses separate neutrino-induced muons from CR muons by selecting only upward going events, since the background from downward going CR muons overwhelms any neutrino-induced muons from above. This reduces the livetime to the time when the Sun is below the Horizon for each experiment. IceCube extended the search in the latest analysis~\cite{Aartsen:2012kia} for the first time to the downward going region by exploiting new atmospheric muon veto techniques, achieving a doubling of the livetime. In order to be sensitive to a wide range of potential WIMP masses, datasets are split into different topological event categories~\cite{Aartsen:2012kia,SuperKWIMPlatest}. Such event categories are classified by contained, partially contained, and through going events, where both contained event classes are dedicated low energy selections. For vertex contained events all initial neutrino energy is deposited inside the detector. Thus, the total reconstructed energy of these events may be used to further discriminate signal from the otherwise irreducible atmospheric neutrino background. Furthermore, particle identification (PI) allows to include electron neutrino event samples in the analysis, doubling the expected signal flux from the Sun. Such an extension of the current searches to include the electron channel is even more rewarding as the atmospheric neutrino background in this channel is much lower (approx. factor 3). These advantages outweigh the drawbacks of a worse angular resolution. PI and precise energy resolution is challenging at sparsely instrumented neutrino telescopes and is consequently only used in Super-K~\cite{SuperKWIMPlatest} thus far. IceCube's low energy extension DeepCore~\cite{Abbasi2012615} or the envisioned low energy in-fill PINGU~\cite{PINGULoI} may also allow for such analysis improvements.
The through going selection provides sensitivity to higher WIMP masses as the neutrino cross-section and the range of the produced muon are proportional to the neutrino energy. Thus, the effective target volume is increased if events
with neutrino interactions outside the instrumented volume are included. Such events can only provide a lower bound on the reconstructed neutrino energy. In this context, spectral information yields less separation power than for contained event samples.  All analyses are generally performed in a blind manner such that the true position of the Sun is not revealed until the selection cuts are finalized.

\begin{figure}[!t]
  \centering
  \includegraphics[width=0.75\textwidth]{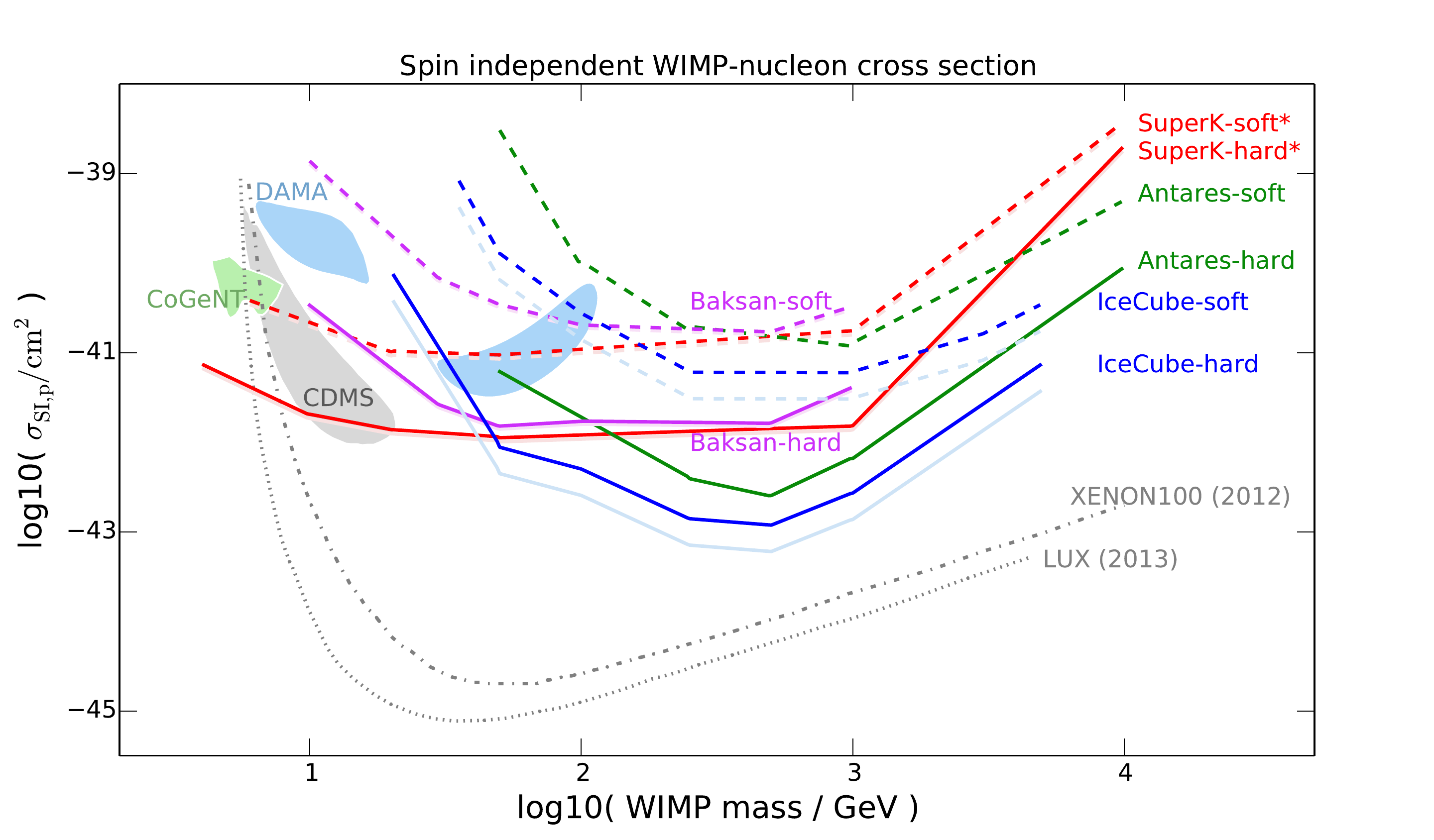}
  \includegraphics[width=0.75\textwidth]{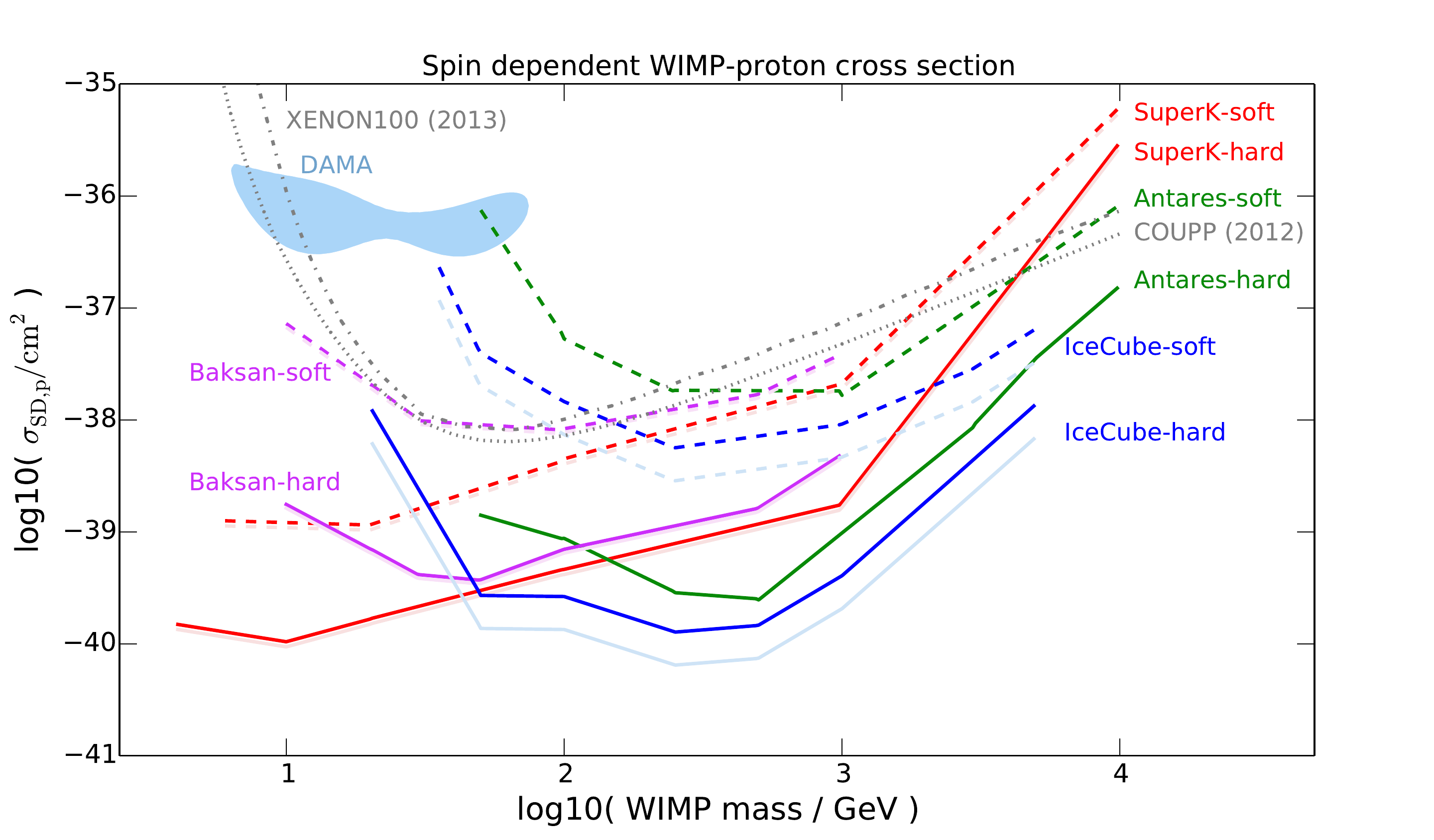}
  \caption{\label{fig:results}  Latest upper limits (90\% CL) on SI (top figure) and SD (bottom figure) WIMP-nucleon cross-sections for hard and soft annihilation channels over a range of WIMP masses from IceCube~\cite{Aartsen:2012kia}, preliminary Super-K~\cite{SuperKWIMPlatest}, ANTARES~\cite{Adrian-Martinez:2013ayv}, and Baksan~\cite{Boliev:2013ai}.  Direct search results from COUPP~\cite{coupp}, XENON100~\cite{xenon,XENON100SD}, preliminary LUX~\cite{lux}, and tentative signal regions~\cite{DAMASavage, COGENTmodulation, CDMSsilicon} are shown for comparison. Expected sensitivities including already recorded data are illustrated with faint lines for each experiment. (SI cross-section results from Super-K are calculated by authors)}
\end{figure}
%

\section{Review of current experimental Searches} \label{current_results}

In this section we give a review of current experimental searches for DM annihilations in the Sun. To this day, all results are consistent with the expected background from atmospheric muons and neutrinos. The upper 90\% confidence level limits on WIMP-nucleon cross-sections for all experiments are shown in Fig.~\ref{fig:results} for SI (top) and SD (bottom) scattering. Additionally, we illustrate the expected sensitivity including already recorded data, under the assumption that all searches are background limited and improve with time, $t$, as $\propto\sqrt{t}$. This simplified calculation allows us to compare the impact of vastly different livetimes on future discovery prospects of the respective detectors. We order the discussion by instrumented detector volume, beginning with the largest detector first. An order-of-magnitude comparison of detector characteristics relevant for current solar DM searches is given in Table~\ref{table:live-time}.\\

IceCube instruments one~km$^{3}$ of glacial ice at the South Pole with 5160~digital optical modules on 86~strings deployed between depths of 1450\,m and 2450\,m. Eight more densely instrumented strings optimized for low energies plus the 12~adjacent standard strings at the centre of the detector geometry make up the DeepCore subarray. IceCube and its predecessor AMANDA previously reported limits on DM annihilations in the Sun with partial detector configurations~\cite{IC22-PRL, PhysRevD.85.042002}. The latest IceCube solar WIMP analysis~\cite{Aartsen:2012kia} uses 317~live-days of data, taken when the detector was operating in its 79-string configuration. For the first time, the DeepCore subarray is included in the analysis, lowering the energy threshold and extending the search to the austral summer (downward going region). The analysis comprises three event selections; summer contained, winter contained, and winter through-going.
As all three data samples are independent, they are combined in one likelihood analysis based on shapes of the space angle distribution with respect to the Sun's position.
The final results are the most stringent SD WIMP-proton cross-section limits to date above 50\,GeV for most WIMP models. Since May 2011, IceCube recorded an additional three years of data in its full 86-string configuration. The projected sensitivity depicted in Fig.~\ref{fig:results} may improve faster than $\sqrt{t}$ in the low mass range below 200\,GeV, as new veto methods against atmospheric muons~\cite{VetoICRCpaper} significantly improve the sensitivities of analyses focused on vertex contained low-energy events.

\begin{table}
  \caption{\label{table:live-time} Rough comparison of neutrino telescope characteristics relevant for current solar DM searches. The median angular resolution ($\overline{\Theta}$) is quoted for different representative neutrino energies (E$_{\nu}$), where applicable. More details in Refs.~\cite{Aartsen:2012kia,MDthesis} (IceCube),~\cite{Adrian-Martinez:2013ayv,AntaresWIMPICRC} (ANTARES),~\cite{SuperKWIMPlatest,Tanaka:2011uf} (Super-K), and~\cite{Boliev:2013ai} (Baksan).}
  \centering
  \begin{tabular}{l l l  l  l  l}
    \hline\hline
    & Datasets with \quad  \quad&  Livetime \quad  \quad&  E$_{\nu}$-range \quad  \quad& Instrumented  \quad \quad& $\overline{\Theta}$ ($^{\circ}$) at E$_{\nu}$   \quad\\
    & completed analyses & (days) & (GeV) & volume (ton) & 25\,/\,100\,/\,1000\,GeV\\\hline
    IceCube & 2010-2011 & 317  & $\gtrsim 10^{\ast}$  &  $\sim$1\,Gton & 13\,/\,3.2\,/\,1.3 \\
    ANTARES$^{\dagger}$ &  2007-2008 & 295 & $\gtrsim 10$   & $\sim$20\,Mton & 6\,/\,3.5\,/\,1.6 \\
    Super-K & 1996-2012 &  3903 &  $\gtrsim 0.1$ & $\sim$50\,kton & 1-1.4$^{\ddagger}$\\
    Baksan & 1979-2009 &  8803 &  $\gtrsim 1^{\ddagger}$ &  $\sim$3\,kton & 1.5$^{\ddagger}$\,(tracks $>$ 7\,m) \\
   \hline\hline 
   \multicolumn{6}{l}{$^{\ast}$ \small{ Threshold corresponds to DeepCore events for this analysis (E$_{\nu} \gtrsim 50$ GeV for non-DeepCore events)~\cite{MDthesis}.}}\\
   \multicolumn{6}{l}{$^{\dagger}$ \small{ Preliminary 2007-2012 results correspond to 1321 days livetime}}\\
   \multicolumn{6}{l}{$^{\ddagger}$ \small{ Values are given at muon level (E$_\mu$); $\overline{\Theta}$ dominated by kinematic scattering angle.}}\\
  \end{tabular}
\end{table}

ANTARES is an undersea neutrino telescope located in the Northern Hemisphere deployed between 2475\,m (seabed) and 2025\,m below the Mediterranean Sea level. The telescope consists of 12 detection lines (450\,m in length) with 25 storeys each (three optical modules per storey). The first results on DM annihilations in the Sun from ANTARES~\cite{Adrian-Martinez:2013ayv} used a dataset recorded with partial detector configurations between January 2007 and December 2008, corresponding to a total livetime of 295 days.
Only upward going events were kept in the analysis, shown in Fig.~\ref{fig:results}. New preliminary results were recently reported~\cite{AntaresWIMPICRC}, including data recorded until 2012 with 1321 live-days total. These limits are comparable to the first results~\cite{ Adrian-Martinez:2013ayv} despite the vast improvement in livetime. Both results can be reconciled, as the first search reported an under fluctuation in the expected background for most tested WIMP signal models. Note, we show no projected sensitivity for ANTARES based on the first results, as they compare to the preliminary new limits.

The Super-K detector is a 50\,kton water Cherenkov detector located at the Kamioka mine in Japan. The inner detector is covered with more than 11100 photomultiplier tubes. Since Super-K started operation in 1996, there have been four experimental phases. The latest preliminary solar WIMP results~\cite{SuperKWIMPlatest} use 3903 days of Super-K \textrm{I}-\textrm{IV} upward going muon data (upmu), which are categorized as the most energetic events in Super-K (see Ref.~\cite{Tanaka:2011uf} for early results). In addition this analysis includes for the first time the contained event class, covering also electron neutrino events, to increase signal acceptance for low mass WIMPs. Results are derived in a combined fit of atmospheric background and WIMP induced neutrinos to data, utilizing angular, spectral and flavour informations. The limits on the SD WIMP-proton cross-section are the most stringent for WIMP masses below 50\,GeV and become significantly weaker at high WIMP masses, due to the small detector volume compared to IceCube or ANTARES, see Fig.~\ref{fig:results}. As Super-K has so far not presented SI cross-section limits, we converted the preliminary SD limits to limits on the SI cross-section, using the method described in Ref.~\cite{Wikstrom:2009kw}, based on latest Super-K data~\cite{SuperKWIMPlatest}. The calculated limits are complementary to direct search results in the very low WIMP mass range, and are in strong tension with alleged WIMP signal interpretations~\cite{DAMASavage, COGENTmodulation, CDMSsilicon}.

The Baksan underground scintillator telescope performs continuous measurements covering 34 years since 1978. Trajectories of penetrating particles are reconstructed using the positions of hit tanks, which represent together a system of 3150 liquid scintillation counters. 
Separation between downward and upward going muons is made by time-of-flight method. The data used for the present analysis~\cite{Boliev:2013ai} have been collected from 1978 till 2009, corresponding to 24.12 years. The Baksan limits on WIMP-nucleon cross-sections (Fig.~\ref{fig:results}) are comparable to those obtained by water Cherenkov detectors for WIMP masses around 100\,GeV, whereas the low and high mass regions are dominated by limits from Super-K and IceCube, respectively.


\section{Astrophysical uncertainties}\label{astro_uncert}

Astrophysical uncertainties on the expected neutrino flux from DM annihilations in the Sun can be grouped in uncertainties on the annihilation rate and those related to neutrino propagation from the core of the Sun to the neutrino detector. Under the assumption of the equilibrium between capture and annihilation of WIMPs in the Sun, the uncertainty on the annihilation rate is given by the uncertainty on the capture rate. In this section we review the uncertainties on the capture process and quantify their impact. We describe qualitatively what changes are expected for the individual uncertainties and invite the reader to use our uncertainty plotter to explore interactively how parameters impact limits and sensitivities. For the reader interested in more background discussion on astrophysical uncertainties we point the reader to Choi.~et.al.~\cite{Choi:2013eda}, which we used as the basis for our interactive tool. As default scenario, denoted by standard model Halo (SMH), we use a local DM density of $\rho_{\chi}$ = 0.3~GeV/cm$^{3}$, a Maxwellian velocity distribution for WIMPs with a 3-D velocity dispersion of 270~km/s with a cut-off at the Galactic escape velocity, and a circular velocity of the Sun of 220~km/s, which are also the defaults in DarkSUSY.

DM gets continuously captured by the Sun and under the assumption of a constant local DM density and an unchanged velocity distribution, the capture rate will be constant. The capture of DM by the Sun will lead to a build up of DM in its core. As the DM density increases in the centre of the Sun, so does the annihilation rate. The annihilation rate scales with the square of the DM density in the Sun and keeps rising up to the point where as much DM is annihilated away as is captured, this is known as the equilibrium condition. If equilibrium is achieved between capture and annihilation then also the uncertainty on the annihilation rate is described by the uncertainty on the capture rate. There is one caveat however, that from the time of capture to thermalization, the time difference should be small on a scale to changing astrophysical conditions. Astrophysical conditions however are predicted to be very stable and a delay in thermalization would even iron out small changes in capture rate, so that the annihilation rate is better described by the time averaged astrophysical conditions. The time to reach equilibrium is known as the equilibration time $\tau$ and typically small compared to the age of the Sun, so that the equilibrium condition is in general very reasonable assumption. For cases where no equilibrium is achieve a scale factor between the capture rate and annihilation rate needs to be used to convert the neutrino flux from the Sun back to capture rate.

The capture rate of DM in the Sun scales linear with the local DM density. Measurements of the local DM density are however not trivial and have resulted in a range of possible values. Commonly used in $\rho_{0}=0.3$~GeV/cm$^3$ as the local DM density at the location of the Sun~\cite{Nakamura:2010zzi},  however, recent measurements favour a higher values in the vicinity of $0.4$~GeV/cm$^3$~\cite{Catena:2009mf,McMillan:2011wd,Salucci:2010qr,Nesti:2013uwa} or even $0.5$~GeV/cm$^3$~\cite{Nesti:2013uwa}. The measurements still have significant uncertainties, so that they overlap,  Gaia data can greatly help to determine the local DM density and distribution~\cite{Bovy:2013raa} and help reduce uncertainties in the measurements. Microlensing and dynamical observations of the Galaxy result in a local DM density of $0.20 - 0.56$~GeV/cm$^3$~\cite{Iocco:2011jz}. Simulations of Milky Way type galaxies, that include baryons, leads to a significant flattening of the DM halo in the direction normal to the stellar disk. Pato et al.~\cite{Pato:2010yq} found that this effect could lead to a DM over density in the local disk of up to 41\% and estimate an average enhancement of 21\% for such a triaxial profile compared to a spherically symmetric Einasto profile, resulting in $\rho_{0} = 0.466 \pm 0.033({\rm stat}) \pm 0.077({\rm syst}) {\rm GeV}/{\rm cm}^3$. Density variation at the position of the Sun are expected to be small and N-body simulation quantify the amount of substructure at the solar circle to be less than 0.1\%~\cite{Springel:2008cc}.

\begin{figure}[!t]
  \centering
  \includegraphics[width=0.85\textwidth]{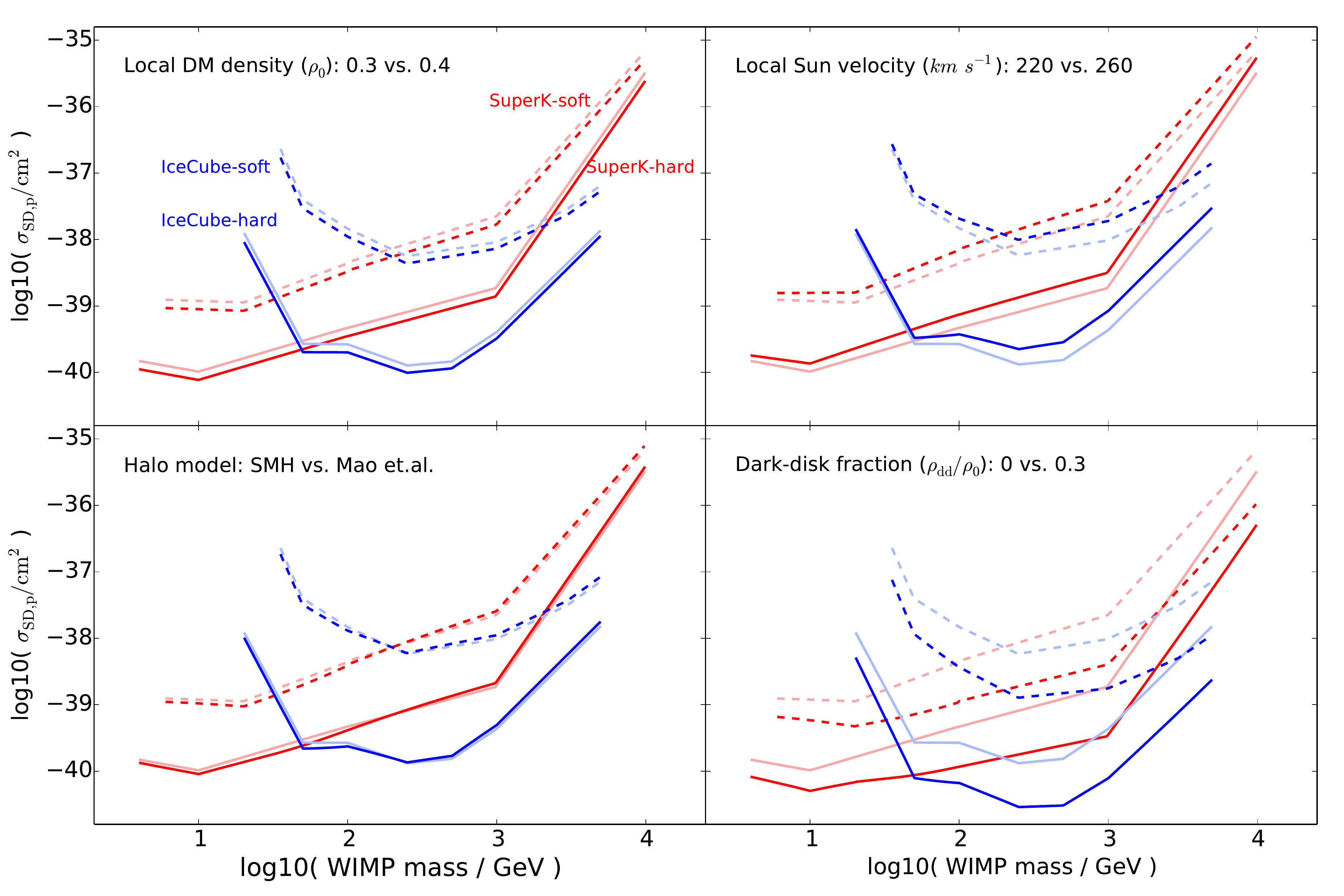}
  \caption{\label{fig:uncertainty} Impact of astrophysical uncertainties on SD cross-section limits from IceCube~\cite{Aartsen:2012kia} and Super-K~\cite{SuperKWIMPlatest} (preliminary). The reference limits including standard assumptions are shown by faint lines (SMH). Respective strong, definite, lines depict impact of varied Astrophysical uncertainty input values: Increase in local DM density (top left), increase in local Sun velocity (top right), different WIMP velocity distribution (bottom left), and additional dark disk fraction (bottom right).}
\end{figure}

The circular velocity of the Sun determines the relative speed with which the Sun plows through the Milky Way WIMP halo and hence enters as astrophysical uncertainty on the capture process. The WIMP halo itself is believed to be non-rotating and under simplest assumptions has a Maxwellian velocity distribution. Measurements of the circular speed of the Sun are intrinsically difficult. Recent estimates based on stellar kinematics~\cite{Schonrich:2012qz} and from maser star forming regions at intermediate radii~\cite{Nesti:2013uwa} yield a central value of around 240~km/s, which is slightly higher than the default value of 220km/s~\cite{Kerr:1986hz}. Other recent measurements, with the exception of  $218\pm 6$~km/s~\cite{Bovy:2012ba}, also result in values of $229\pm18$~km/s~\cite{Ghez:2008ms}, $254\pm16$~km/s~\cite{Reid:2009nj}, and $244\pm 13$~km/s~\cite{Bovy:2009dr}. A change of circular velocity by 10\%, change the capture rate by up to about 25\%~\cite{Choi:2013eda} for both the SI and the SD case. For WIMP masses below 100~GeV the effect is considerably smaller (see Figure~\ref{fig:uncertainty} as an example).

The velocity distribution of WIMPs in the Galactic Milky Way halo is assumed to be Maxwellian as would be expected from particles in a self-gravitating DM halo under thermal equilibrium. The high-velocity tail of the distribution is expected to be cut-off at the Galactic escape velocity, as particles would just leave the halo. The high-velocity cut-off has only a very minor impact on capture rates. It only becomes relevant for WIMP masses below about 20~GeV and even for those changes capture rates by less than 3\%. N-body simulations, with and without baryons, both confirm that WIMP velocity distributions resemble Maxwellian distributions, but can have some substantial overall structural differences, which can lead to a rather complex change in capture rates as function of the WIMP mass. The capture behaviour can be understood if we remind ourselves that WIMPs can only be captured if they fall below the escape velocity of the Sun. If a WIMP is heavy, the relative velocity needs to be smaller to loose the same momentum in a single scatter compared to a lighter one. Figures~1 and~2 of Choi et al.~\cite{Choi:2013eda}, show this behaviour and it can also be interactively explored via our plotter and visualized as an example in Fig.~\ref{fig:uncertainty}. We show the change in limits based on a standard Maxwellian halo and three different distributions obtained from various simulations~\cite{Vogelsberger:2008qb,Ling:2009eh,Kuhlen:2009vh,Mao:2012hf}. The velocity distribution functions of Vogelsberger et al~\cite{Vogelsberger:2008qb}, Ling et. al.~\cite{Ling:2009eh}, and Mao et al.~\cite{Mao:2012hf} originate respectively from the Aquarius~\cite{Springel:2008cc} project, which resolved a Milky Way-sized galactic halo with more than a billion particles; an N-body simulation with Baryons~\cite{Ling:2009eh} carried with the cosmological Adaptive Mesh Refinement code RAMSES~\cite{Teyssier:2001cp}; the Rhapsody cluster re-simulation project~\cite{Wu:2012wu}.

While the various velocity distributions of a self-gravitating halo have a relatively small impact on the capture rate, a co-rotating dark disk can have a significantly larger impact. The Milky Way merger history favours the existence of such a co-rotating structure created from materials accreted from satellites galaxies. Simulations show that the local density of the dark disc could range from a few percent~\cite{Kuhlen:2013tra} up to the same magnitude as the local DM halo~\cite{Read:2009iv}. If the disk is co-rotating, the relative velocity between the DM in the disk and the Sun is small and hence easy to capture. Capture rates for 100~GeV WIMPs can be increased by a factor of  12 and 5 for SI and SD processes, respectively~\cite{Choi:2013eda,Bruch:2008rx,Kuhlen:2013tra,Purcell:2009yp,Frandsen:2011gi,Billard:2012qu}. Figure~\ref{fig:uncertainty} shows the impact of the dark disk and that indirect limits could drastically be increased compared to direct searches, that have almost no enhancement in rates due to a dark disk~\cite{Ling:2009cn}.

The phase space of WIMPs could be influenced by the gravitational influence of the major planets. After extensive discussions~\cite{Gould:1990ad,Gould:1999je,Damour:1998rh,Lundberg:2004dn,Peter:2009mi,Peter:2009mk, Peter:2009mm,Sivertsson:2009nx} the latest study~\cite{Sivertsson:2012qj} however concluded that one can to high precision use Liouville's theorem for weakly captured WIMPs, not just for the gravitationally captured WIMPs as previously believed. The solar weak capture process is hence to be viewed as an extra source of WIMP diffusion in the solar System, and that effects of planets can be ignored.

Another source of uncertainty is the elemental and structural composition of the Sun and nuclear form factors (not included in interactive tool at present). 
For calculations of $\sigma_{\mathrm{SD}}$ the nuclear form factor uncertainty is negligible, because the capture is dominated by protons. For $\sigma_{\mathrm{SI}}$ interactions heavy elements are important in the capture process. Assuming various models of the form factor, the uncertainty on $\sigma_{\mathrm{SI}}$ can be as much as 20\%~\cite{PaoloNuclearFormFactor,Ellis:2008hf,Bottino:1999ei,deAustri:2013saa}. The dependence of the capture rate on the abundances of elements in the Sun is evaluated by comparing different composition models. An uncertainty of at most 4\% difference in the annihilation rates between the models based upon the two most recent abundance estimates (AGSS09 and AGSS09ph) is calculated in Ref.~\cite{Ellis:2009ka}. The authors of Ref.~\cite{Ellis:2009ka} conclude that the discrepancy between meteoritic and photospheric measurements is not a significant issue in estimating annihilation rates.

Very light DM could also become unbound from the Sun in a process known as evaporation, this effect is only relevant for WIMP masses below about 4~GeV~\cite{Griest:1986yu,Gould:1987ju,1990ApJ...356..302G,Bernal:2012qh,Kappl:2011kz}. This mass range is not the focus of this review and evaporation has hence been ignored.


\section{Conclusions and Potential for solar WIMP detection in the future}\label{conclusion}

Three strategies are being pursued to determine the nature of DM. Colliders recreate the energy densities that existed in the early universe, when DM was thermally produced and direct and indirect searches try to see signatures based on the remnant DM in the universe from this initial production. While all approaches will be needed to eventually gain a more complete picture about the underlying theory of DM and its role in the universe, solar WIMPs have a unique place in all these approaches. Solar WIMP searches combine high discovery potential through an unmistaken signature with dependencies on the DM halo conditions distinctively different from direct searches. We have reviewed the DM capture process in the Sun, discussed relevant astrophysical uncertainties, and current searches for solar WIMPs. A novelty with our review is an interactive plotting tool that allows the user to explore current limits and impacts from astrophysical uncertainties. 

Current limits from solar WIMP searches already exclude many of the DM hints seen in direct detection experiments, even under more extreme WIMP halo conditions as our interactive plotting tool can also demonstrate. While DM has yet evaded detection via signals from the Sun, the prospects for a discovery are not slim given possible improvements in analyses, more data and new detectors. Classical searches using the neutrinos generated in the decay of annihilation products have achieved very high efficiencies and currently we expect IceCube and ANTARES to improve with respect to Super-K and Baksan, due to the accumulation of lifetime in these young experiments. As IceCube already provides the best sensitivity for high WIMP masses, this search continues to hold a discovery potential. The latest IceCube analysis also only used the partially completed DeepCore detector, so that up-coming analyses will also achieve sensitivities to slightly lower WIMP masses.

Super-K's preliminary results already combine different energy dependent event topologies, as well as neutrino flavours. Inclusion of similar technique for IceCube and ANTARES could yield also some improvement in sensitivity.  

A reliable tau neutrino identification could also post a discovery potential for solar WIMP search, as these are background free in down-going events. Various papers have discussed this possibility~\cite{Fornengo:2011em}, however the smaller cross-section as well as the experimental challenge to reliable identify tau events makes this a very difficult undertaking. Large detectors with Liquid Argon Time Projection Chambers could offer a possibility to tag tau events with a relatively small background~\cite{Conrad:2010mh}. It is however difficult to see how these detectors could compete with very large volume water and ice Cherenkov detectors such as Hyper-K and PINGU. PINGU has already evaluated their solar WIMP sensitivities and are provided with the plotter.

Searches for DM with liquid scintillator detectors, like KamLAND, or in the future with JUNO or RENO-50 also offer interesting prospects~\cite{Kumar:2011hi}.

Neutrino detectors with an energy threshold of about 20~MeV could also look for solar WIMPs using neutrinos from stopped pion decays from hadronic showers in the Sun. This signal has a very weak dependence on the mix of annihilation channels and hence the model. The weak dependence on the mix of annihilation channels is a result of the fact that most annihilation final states will eventually decay or produce hadrons giving rise to this low-energy neutrino signal. Further, one could envision scenarios in which there are very few or no high-energy neutrinos produced, for example if annihilation final states are predominantly light quarks. For such a scenario or for the case that DM annihilates to electron-positrons, which then create hadrons in inelastic collisions in the Sun, a low-energy neutrino signal would be present from pion decay at rest~\cite{Bernal:2012qh,Rott:2012qb}.


\section{Acknowledgments}
CR acknowledges support from the Basic Science Research Program through the National Research Foundation of Korea funded by the Ministry of Education, Science and Technology (2013R1A1A1007068). We would like to thank Koun Choi for providing DarkSUSY generated WIMP capture tables. We thank the referees for comments and suggestions that helped us to improve this review.


\appendix

\section{Astrophysical Uncertainty interactive tool}\label{toolDescription}

\begin{figure}[!t]
  \centering
  \includegraphics[width=0.9\textwidth]{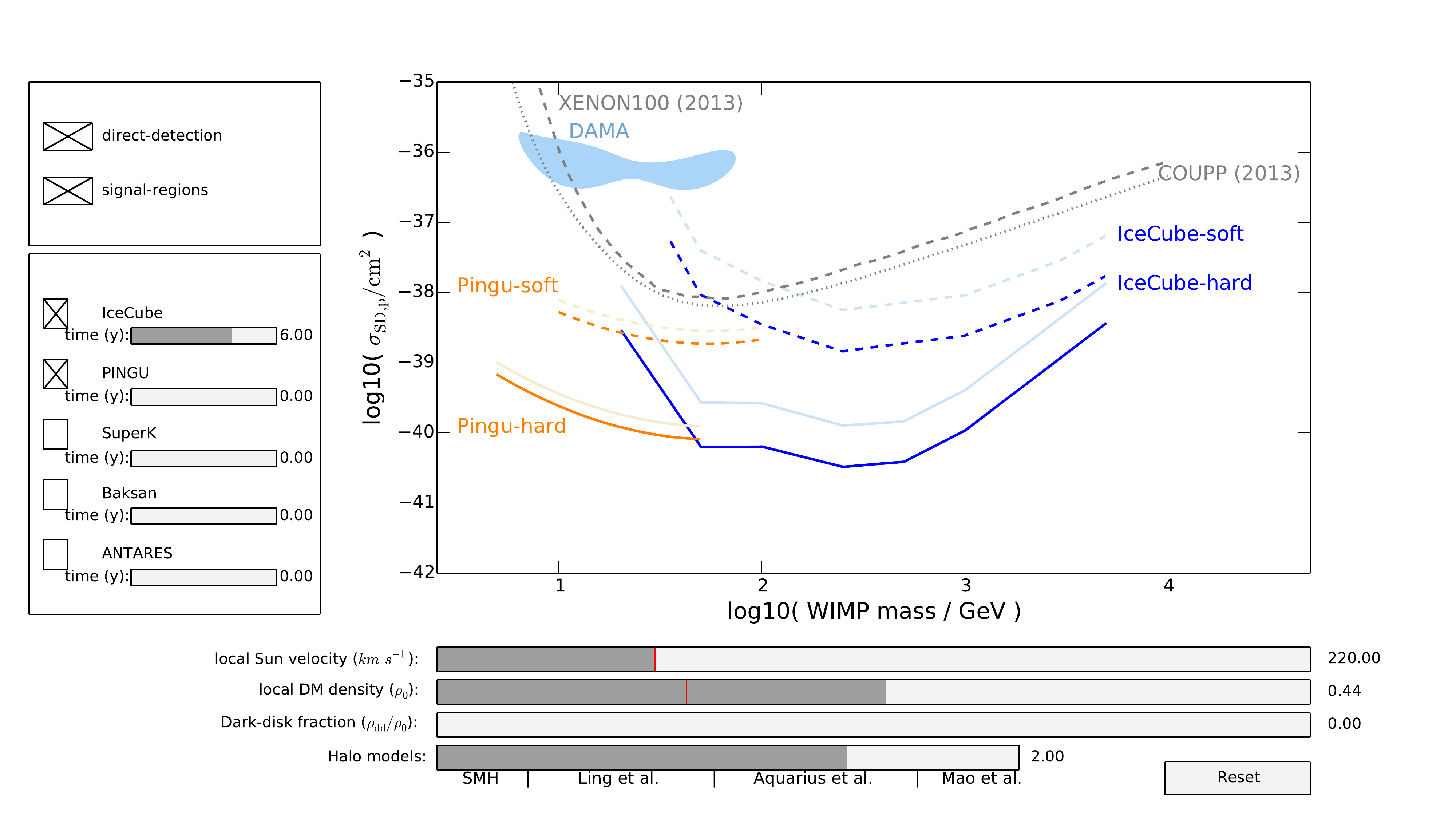}
  \caption{\label{fig:screenshot} Example screenshot of a possible configuration of the interactive plot for SD cross-section results (see text for detailed description of options and displayed parameter values). The figure shows limits from IceCube~\cite{Aartsen:2012kia} and the projected sensitivity from PINGU~\cite{PINGULoI}, as well as direct search results from COUPP~\cite{coupp}, XENON100~\cite{XENON100SD}, and a tentative signal region~\cite{DAMASavage}. The tool provides one slider per experiment to scale current limits in time, as well as three continuos sliders for uncertainties on the local Sun velocity, $\rho_{0}$, and the dark disk fraction. The fourth slider allows to visualize the impact of four different Halo models on cross-section limits. The box on the left refers to additional years of data, compared to the current dataset or in the case of PINGU to one year of data for the sensitivity. }
\end{figure}

In this appendix we give a detailed description of the interactive plotting tool, which incorporates most sources of Astrophysical uncertainties and experimental constraints discussed in sections~\ref{current_results} and~\ref{astro_uncert}. This comprehensive interactive tool aims to quickly visualize the impact of various uncertainties on experimental limits, see example screenshot in Fig.~\ref{fig:screenshot}. We list all available options and detail what was done to develop this tool. In addition, we quantify the secondary uncertainties introduced with the fitting method.\\

We developed one interactive plot for SD and SI cross-section results each, based on scripts using python and Matplotlib\footnote{Minimum requirements: Matplotlib version-1.3.0, numpy version-1.5, and python version-2.5}~\cite{Hunter:2007}. Figure~\ref{fig:screenshot} shows a screenshot of a typical SD plot. There are four solar DM experimental results and one future projection to select from: The latest IceCube, Super-K, Baksan, and ANTARES results, as well as the projected PINGU sensitivity. Additionally, these constraints can be compared to several direct detection limits (LUX~\cite{lux}, XENON100~\cite{XENON100SD}, and COUPP~\cite{coupp}) and tentative WIMP signal interpretations in the the low WIMP mass region~\cite{DAMASavage, COGENTmodulation, CDMSsilicon}. 
The selected list of shown limits from direct detection experiments is composed with a focus on the most recent and stringent results for a wide range of viable WIMP masses\footnote{Note, limits from PICASSO~\cite{picasso} and SIMPLE~\cite{simple} (SD cross-section), and CDMS~\cite{cdms} and CoGeNT~\cite{cogent} (SI cross-section) are more stringent for very low WIMP masses than the selected results.} .
As discussed in section~\ref{current_results}, all results from neutrino telescopes use datasets with largely different livetimes. In this context, we included a slide bar allowing to adjust the additional lifetime in years for every experiment separately. We included four Astrophysical uncertainties in the plot, where characteristic values can be varied within an expected range, and their impact on cross-section limits immediately visualized. The sliders for the uncertainty in the solar circular velocity, local DM density ($\rho_{0}$), and dark disk fraction ($\rho_{\rm dd}/\rho_{0}$) are continuous. The corresponding input parameters may be altered in the range from 200 to 280\,${\rm kms}^{-1}$, 0.1 to 0.8\,${\rm GeV}{\rm cm}^{-3}$, and 0 to 1, respectively. The slider for different Halo models is discrete, where the valid ranges for each model is indicated. Default values (start values) are marked in red, and correspond to a SMH. The reference limit (default input) is always fixed and depicted by faint lines for each experiment in the respective line colour. This allows for better comparison between reference limits and such for user specific input. The `Reset' button restores all default values. The example (Fig.~\ref{fig:screenshot}) shows SD cross-section limits for direct searches, low WIMP mass signal regions, IceCube, and the projected 1 year PINGU sensitivity. PINGU is currently envisioned and may be deployed within the next five years. We anticipate these projected future limits by scaling current IceCube limits with additional six years of data. Furthermore, we selected for the example default values for the solar circular velocity and dark disk fraction, but values of $\rho_{0}$=0.44 and an Halo model from~\cite{Vogelsberger:2008qb}. 

In section~\ref{astro_uncert}, we reviewed the impact of Astrophysical uncertainties on solar DM searches, and concluded that resulting uncertainties on the annihilation rate can be parameterized by the uncertainty on the capture rate. In order to allow for interactive manipulation of input parameters, experimental limits and Astrophysical uncertainties have to be available in functional form. We use spline interpolation to achieve sufficiently smooth piecewise defined functions. Experimental results are obtained with linear interpolation between data points. The authors of Ref.~\cite{Choi:2013eda} parameterize Astrophysical uncertainties by WIMP mass and associated characteristic value, i.e. the local Sun velocity for the case of solar circular velocity uncertainties. These more complex dependencies are fitted with higher order polynomial functions in two dimensions. The fitting process introduces new secondary uncertainties. These are quantified by calculating the variance of the distribution of ratios $x_{i}/f(x_{i})$ for each fit separately, where $x_{i}$ is the data point at position $i$ and $f(x_{i})$ the fit-function evaluated at $x_{i}$. The mean values of these distributions are consistent with 1 and corresponding standard deviations are listed in table~\ref{table:fit-uncert}.\\
\begin{table}
  \caption{\label{table:fit-uncert} Summary of fit uncertainties given at 1~sigma level.}
  \centering
  \begin{tabular}{l | c  c}
    \hline\hline
    fit& SD cross-section & SI cross-section\\\hline
    experimental data & 0.012 &  0.013  \\ 
    solar circular velocity  & 0.011 &   0.012  \\
    dark disk fraction   & 0.062 &   0.061  \\ 
    Halo model  & 0.012 &   0.009  \\  
    \hline\hline
  \end{tabular}
\end{table}

In a concluding remark, we want to point out that one viable future extension to the interactive tool may be the inclusion of Astrophysical uncertainties in direct detection experiments. The default scenario (SMH), defined in section~\ref{astro_uncert}, is currently assumed for all direct detection limits.\\

\bibliographystyle{model1-num-names}

\hyphenation{Post-Script Sprin-ger}

\end{document}